# Low-Dose CT via Deep Neural Network

Hu Chen[1, 2], Yi Zhang[1, *], Weihua Zhang[1], Peixi Liao[3], Ke Li[1, 2], Jiliu Zhou[1], Ge Wang[4]

1. College of Computer Science, Sichuan University, Chengdu 610065, China
2. National Key Laboratory of Fundamental Science on Synthetic Vision, Sichuan University, Chengdu 610065, China
3. Department of Scientific Research and Education, The Sixth People's Hospital of Chengdu, Chengdu 610065, China
4. Department of Biomedical Engineering, Rensselaer Polytechnic Institute, Troy, NY 12180, USA
* Corresponding author: yzhang@scu.edu.cn

**Abstract**: In order to reduce the potential radiation risk, low-dose CT has attracted more and more attention. However, simply lowering the radiation dose will significantly degrade the imaging quality. In this paper, we propose a noise reduction method for low-dose CT via deep learning without accessing the original projection data. An architecture of deep convolutional neural network was considered to map the low-dose CT images into its corresponding normal-dose CT images patch by patch. Qualitative and quantitative evaluations demonstrate a state-the-art performance of the proposed method.
**Key words**: low-dose CT, noise reduction, deep learning, convolutional neural network

## 1. Introduction

In recent decades, X-ray computed tomography has been widely used in both diagnostic and industrial fields. With the increasing number of CT scans, the potential radiation risk attracts an increasingly more public concern [1, 2]. Most current commercial CT scanners utilize the filtered backprojection (FBP) method to analytically reconstruct images. One of the most used methods to reduce the radiation dose is to lower the operating current of the X-ray tube. However, directly lowering the current will significantly degrade the imaging quality due to the excessive quantum noise caused by an insufficient number of photons in the projection domain.

Many approaches were proposed to improve the quality of low-dose CT images. These approaches can be categorized into three classes: sinogram filtering, iterative reconstruction and image processing.

Sinogram filtering directly smoothens raw data before FBP is applied. Structural adaptive filtering and bilateral filtering are two efficient methods proposed by Balda's and

Manduca's groups respectively [3, 4]. After investigating the statistical model of noisy sinogram data, Li et al. developed a penalized likelihood method to suppress the quantum noise [5]. Two groups respectively improved this method with multiscale decomposition [6, 7]. Iterative reconstruction solves the problem iteratively, aided by some kinds of prior information on target images. Different priors were proposed, including total variation (TV) [8-11], nonlocal means (NLM) [12, 13], dictionary learning [14] and low rank matrix decomposition [15]. Despite the successes achieved by these two approaches, they are often restricted in practice due to the difficulty of gaining well-formatted projection data since the vendors are not generally open in this domain. Meanwhile, the iterative reconstruction methods involve heavy computational costs. In contrast to the first two categories, image processing does not rely on the projection data, can be directly applied on low-dose CT images, and easily integrated into the current CT workflow. However, it is underlined that the noise in low-dose CT images does not obey a uniform distribution. As a result, it is not easy to remove image noise and artifacts completely with traditional image denoising methods. Extensive efforts were made to suppress image noise via image processing for low-dose CT. Chen et al. and Ma et al. introduced NLM for low-dose CT image restoration with similar means [16, 17]. Li et al. improved this measure with a local noise level estimation [18]. Based on the popular idea of sparse representation, Chen et al adapted K-SVD [19] to deal with low-dose CT images [20]. Also, block-matching 3D (BM3D) algorithm has been proved powerful in image restoration for different noise types and several CT imaging tasks [21-23].

Recently, deep learning (DL) has generated an excitement in the field of machine learning and computer vision. DL can efficiently learn high-level features from the pixel level through a hierarchical multilayer framework [24-26]. Several network architectures were proposed which produced promising results for image restoration. Jain and Seung first used the convolutional neural networks (CNN) architecture for image restoration in which an unsupervised learning procedure was used to synthesize clean images [27]. A similar network architecture was tested for image super-resolution [28] and deconvolution [29]. Since the denoising autoencoder (DA) is a natural tool to improve noisy input samples, Xie et al. constructed a deep network with stacked sparse denoising autoencoder (SSDA) and demonstrated its denoising and inpainting ability [30]. In [31], the performance of a pre-training plain multi-layer perceptron (MLP) was evaluated against the current state-of-the-art image denoising methods. Following-up these studies, several variants were proposed [32, 33]. In the field of medical image processing, there are already multiple papers on DL-based image analysis, such as image segmentation [34-36], nuclei detection [37, 38], organ classification [39].

To our best knowledge, however, there are few studies proposed for imaging problems. In this regard, Wang et al introduced the DL-based data fidelity into the framework of iterative reconstruction for undersampled MRI reconstruction [40]. Zhang et al. proposed a limited-angle tomography method with deep CNN [41]. Wang shared his opinions on deep learning for image reconstruction [42].

Inspired by the great potential of deep learning in image processing, here we propose a deep convolutional neural network to map low-dose CT images into corresponding normal-dose CT images. An offline training stage is needed using a reasonably sized training set. In the second section, the network details are described. In the third section, qualitative and quantitative results are presented. Finally, in the last section, relevant issues are discussed, and the conclusion is drawn.

## 2. Methods

### 2.1 Noise reduction model

Due to the encryption of the raw projection data, post-reconstruction restoration is a reasonable alternative for sinogram-based methods. Once the target image is reconstructed from a low-dose scan, the problem becomes image restoration. The only difference between low-dose CT and natural image restoration is that the statistical property of low-dose CT images cannot be easily determined in the image domain. This property will significantly compromise the performance of noise-dependent methods, such as median filtering, Gaussian filtering, anisotropic diffusion, etc., which were respectively designed for specific noise types. However, learning-based methods are immune to this problem, because this kind of methods is strongly dependent on training samples, instead of noise type. We model the noise reduction problem for low-dose CT as follows.

Let $\mathbf{X} \in \mathbb{R}^{m \times n}$ is a low-dose CT image and $\mathbf{Y} \in \mathbb{R}^{m \times n}$ is the corresponding normal-dose image, then a relationship can be formulated as:

$$\mathbf{X} = \sigma(\mathbf{Y}), \tag{1}$$

where $\sigma: \mathbb{R}^{m \times n} \to \mathbb{R}^{m \times n}$ represents the corrupting process due to the quantum noise that contaminates the normal-dose CT image. Then, noise reduction problem can be converted to find a function $f$:

$$f = \arg\min_{f} \| f(\mathbf{X}) - \mathbf{Y} \|_2^2, \tag{2}$$

where $f$ is treated as the best approximation of $\sigma^{-1}$.

**2.2 Convolutional Neural Network**

In this study, the low-dose CT problem is solved in three steps: patch coding, non-linear filtering, and reconstruction. Now. we introduce each steps in details.

*2.2.1 Patch encoding*

Sparse representation (SR) has been popular in the field of image processing. The key idea of SR is to represent extracted patches of an image with a pre-trained dictionary. Such dictionaries can be categorized into two groups according to how dictionary atoms are constructed. The first group is the analytic dictionary such as DCT, Wavelet, FFT, etc. [43]. The other one is learned dictionary [44], which has better adaptivity and can preserve more application-specific details assuming proper training samples. This method can be treated as convolution operations with a series of filters, each of which is an atom. Our method is similar to the latter one, and SR is involved in the step for patch encoding in the form of optimizing a neural network. First, we extract patches from training images with a fixed slide size. Second, the first layer to implement patch coding can be formulated to:

$$C_1(\mathbf{y}) = \mathrm{ReLU}(\mathbf{W}_1 * \mathbf{y} + \mathbf{b}_1), \tag{3}$$

where $\mathbf{W}_1$ and $\mathbf{b}_1$ denote the weights and biases respectively, $*$ represents the convolution operator, $\mathbf{y}$ is extracted patch from images, and $\mathrm{ReLU}(x) = \max(0, x)$ is the activation function [45]. In CNN, $\mathbf{W}_1$ can also be seen as $n_1$ convolution kernels with a uniform size of $s_1 \times s_1$. After patch encoding, we embed the image patches into a feature space, and the output $C_1(\mathbf{y})$ is a feature vector, whose size is $n_1$. Correspondingly, $\mathbf{b}_1$ has the same size as that of $\mathbf{W}_1 * E(\mathbf{y})$.

*2.2.2 Non-linear filtering*

After processed by the first layer, a $n_1$-dimensional feature vector is obtained from the extracted patch. In the second layer, we transform these $n_1$-dimensional vectors into $n_2$-dimensional ones. This operation is equivalent to a filtration on the feature map from the first layer. The second layer to implement non-linear filtering can be formulated as:

$$C_2(\mathbf{y}) = \text{ReLU}(\mathbf{W}_2 * C_1(\mathbf{y}) + \mathbf{b}_2), \tag{4}$$

where $\mathbf{W}_2$ is composed of $n_2$ convolution kernels with a uniform size of $s_2 \times s_2$ and $\mathbf{b}_2$ has the same size as that of $\mathbf{W}_2 * C_1(\mathbf{y})$. If the desired network only has two layers, the output $n_2$-dimensional vectors of this layer are the corresponding cleaned patches for the final reconstruction. Meanwhile, inserting more layers is a possible way to potentially boost the capacity of the network. However, a deeper CNN is at a cost of more complex computation especially longer training time.

*2.2.3 Reconstruction*

In this step, the processed overlapping patches are merged into the final complete image. These overlapping patches are properly weighted before their summation. This operation can also be considered as filtration by a pre-defined convolutional kernel, as formulated by

$$C(\mathbf{y}) = \mathbf{W}_3 * C_2(\mathbf{y}) + \mathbf{b}_3, \tag{5}$$

where $\mathbf{W}_3$ is composed of only 1 convolution kernel with a size of $s_3 \times s_3$, and $\mathbf{b}_3$ has the same size as that of $\mathbf{W}_3 * C_2(\mathbf{y})$.

Eqs. (3)-(5) are all convolutional operations, although they have been designed for different purposes. That is the reason why the CNN architecture is in use for our low-dose CT image denoising.

*2.2.4 Training*

Once the network is configured, the parameter set, $\Theta = \{\mathbf{W}_1, \mathbf{W}_2, \mathbf{W}_3, \mathbf{b}_1, \mathbf{b}_2, \mathbf{b}_3\}$, of the network must be estimated to learn the function $C$. Given the training dataset $D = \{(\mathbf{x}_1, \mathbf{y}_1), (\mathbf{x}_2, \mathbf{y}_2), \ldots, (\mathbf{x}_N, \mathbf{y}_N)\}$ with $\{\mathbf{x}_i\}$ and $\{\mathbf{y}_i\}$ which respectively denote normal-dose image patches and its corresponding noisy versions, and $N$ is the total number of training samples, the estimation of the parameters can be achieved by minimizing the following loss function in terms of the mean squared error:

$$L(D;\Theta) = \frac{1}{N}\sum_{i=1}^{N}\|\mathbf{x}_i - C(\mathbf{y}_i)\|^2. \tag{6}$$

The loss function is optimized using the stochastic gradient descent method [46].

## 3. Experiment

There are many factors that may affect the performance of the proposed model, here we focus on key ones including dose setting, training data and testing data. Meanwhile, state-of-the-arts methods, both iterative reconstruction and post-reconstruction restoration, were selected for comparison, including:

1) ASD-POCS [8]: This method is an early iterative CT reconstruction method with TV being used as an efficient sparse constraint.
2) K-SVD [20]: This is a typical dictionary learning based method. The noisy patch can be restored by being approximately represented as a sparse linear combination of learned dictionary atoms.
3) BM3D [22]: This is currently the most popular denoising method, which is based on the self-similarity of images.

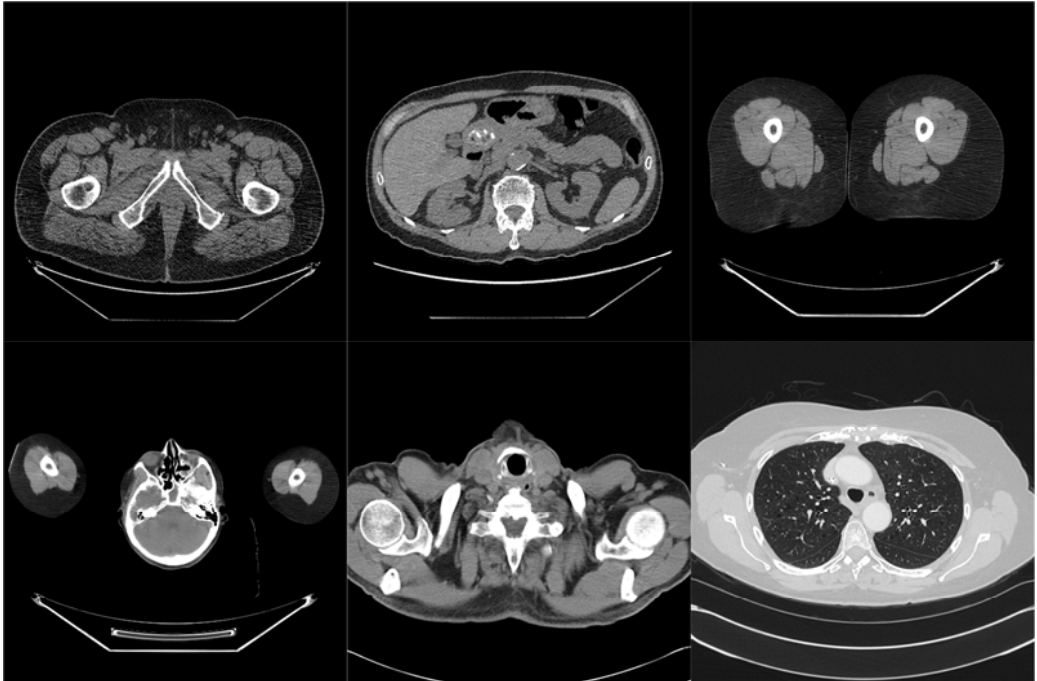

Fig. 1. Typical CT images involved in the training set.

The parameters for the comparative methods were set according to the recommendation in the original references.

The peak signal to noise ratio (PSNR), root mean square error (RMSE) and structure similarity index (SSIM) [47] were used as quantitative metrics. All the experiments were tested using MATLAB 2015b on a PC (Intel i7 6700K CPU, 16 GB RAM and GTX 980 Ti graphics card).

### 3.1 Dataset preparation

7,015 CT normal-dose images with size of 256×256 from 165 patients including different parts of human body were downloaded from NCIA (National Cancer Imaging Archive). Fig. 1 illustrates several typical slices included in our training set.

The corresponding low-dose images were generated by imposing Poisson noise into each detector element of the simulated normal-dose sinogram with the blank scan flux $b_0 = 10^5$. Siddon's ray-driven algorithm [48] was used to simulate fan-beam geometry. The source-to-rotation center distance was 40 cm while the detector-to-rotation center was 40 cm. The image array was 20 cm × 20 cm. The detector width was 41.3 cm in length containing 512 detector elements. The data were uniformly sampled in 1024 views over a full scan. The input patches of the network were extracted from the original images with size $m = 33$. The slide size was 1. The original 200 training images included in our first experiment resulted in about $10^6$ samples. There are two reasons why patches were used, instead of whole images: one is that the images can be better represented by local structures; the other is that deep learning requires a big training dataset and chopping the original images into patches can efficiently boost the number of samples.

For fairness, in the testing stage, 100 low-dose images were randomly selected from the original dataset as the testing set, and 200 images excluding the ones from the same patients involved in testing set were randomly selected as training set.

### 3.2 Parameter setting

In this paper, three layers were used in the proposed network. The filter number, $n_1$ and $n_2$, were respectively set to 64 and 32, and the corresponding filter sizes, $s_1$, $s_2$ and $s_3$, were set to 9, 3 and 5. The initial weights of filters in each layer were randomly set, which satisfies the Gaussian distribution with zero mean and standard deviation 0.001. The initial learning rate was 0.001 and slowly decayed to 0.0001 during the training process.

**3.3 Results**

*3.3.1 Visual inspection*

In this section, we select 2 representative slices, which contain different parts (chest and abdomen) of human body from the results of testing set. Fig. 2 and 3 gives the results from different methods.

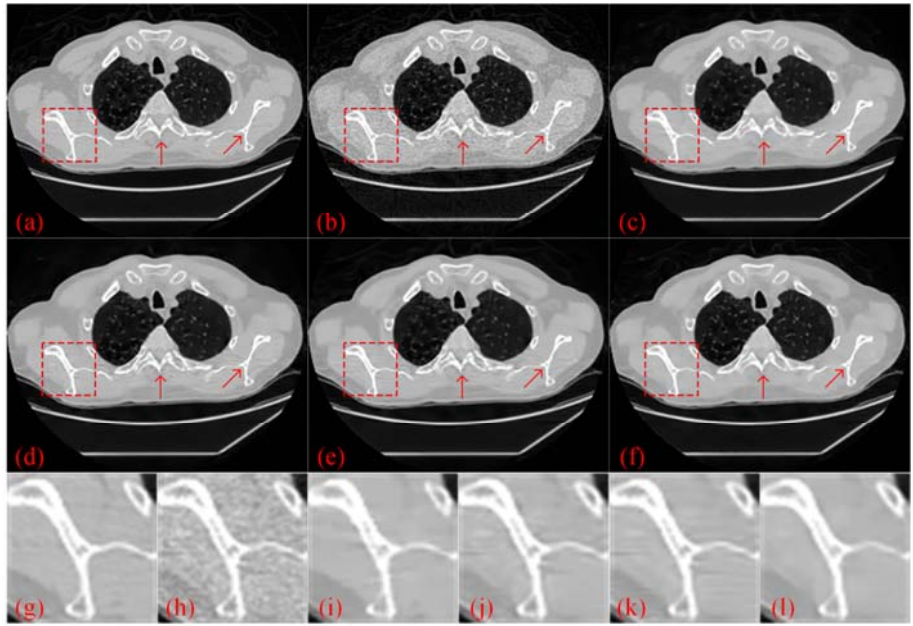

Fig. 2. Results of a slice of chest images. (a) Original normal-dose image; (b) the low-dose images; (c) the ASD-POCS reconstructed image; (d) the KSVD processed low-dose image; (e) the BM3D processed low-dose image; (f) the CNN processed low-dose image; (g)-(l) demonstrate the zoomed regions specified with red box in (a)-(f).

In both Fig. 2 and 3, the noise and artifacts caused by the lack of incident photons severely degrade the imaging quality. Some details and important structures cannot be discriminated. All of the methods can eliminate the noise and artifacts in different degrees. However, due to the piecewise constant assumption of TV, ASD-POCS causes blocky effects in both Fig. 2 and 3. KSVD and BM3D cannot efficiently suppress the streak artifacts near the bone as indicated by red arrows in Fig. 2 and the zoomed parts (Fig.2 (g)-(l)) from red boxes can show more details. In Fig.3, the arrows point to small details and boundaries and only the proposed CNN based method can recover most of these details.

*3.3.2 Quantitative measurement*

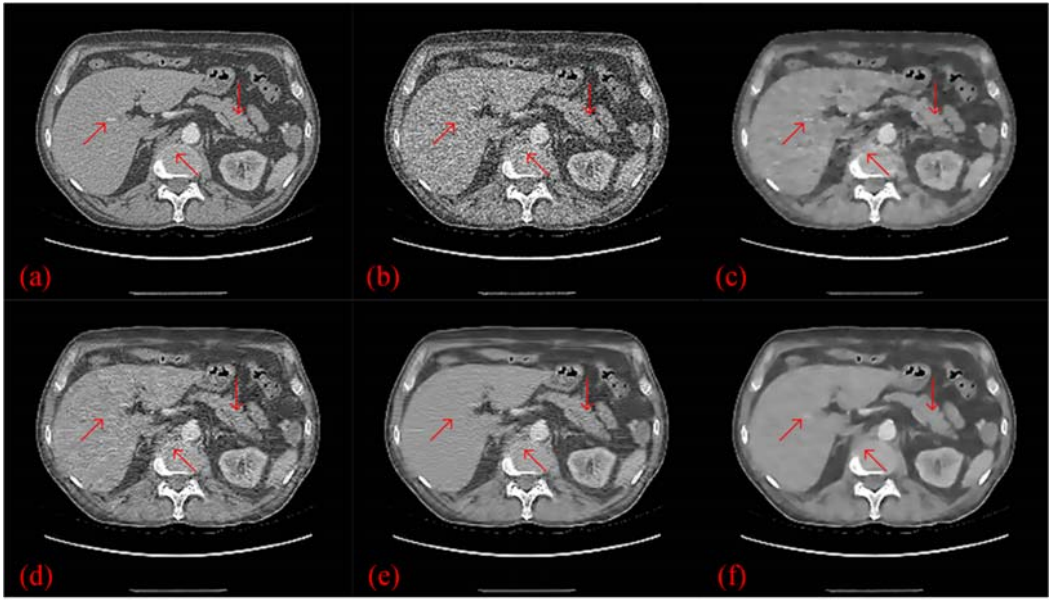

Fig. 3. Results of a slice of abdomen images. (a) Original normal-dose image; (b) the low-dose images; (c) the ASD-POCS reconstructed image; (d) the KSVD processed low-dose image; (e) the BM3D processed low-dose image; (f) the CNN processed low-dose image.

Table I Quantitative measurements associated with different algorithms for the image in Fig.2

|      | Low-dose | TV    | KSVD   | BM3D   | CNN200  |
|------|----------|-------|--------|--------|---------|
| PSNR | 37.0788  | 41.22 | 40.6901| 40.9998| **41.6843** |
| RMSE | 0.014    | 0.0087| 0.0092 | 0.0089 | **0.0082**  |
| SSIM | 0.9026   | 0.9527| 0.9592 | 0.9657 | **0.9721**  |

Table II Quantitative measurements associated with different algorithms for the image in Fig.3

|      | Low-dose | TV     | KSVD   | BM3D   | CNN200  |
|------|----------|--------|--------|--------|---------|
| PSNR | 34.3094  | 37.5485| 38.3841| 38.9903| **38.9904** |
| RMSE | 0.0193   | 0.0133 | 0.012  | 0.0112 | **0.0101**  |
| SSIM | 0.8276   | 0.8825 | 0.9226 | **0.9295** | 0.9277 |

To quantitatively evaluate the proposed CNN based algorithm, PSNR, RMSE and SSIM are measured for all the images in testing set. The results for the restored images in Fig. 2 and 3 are listed in Table I and II.

It can be seen in Table I that, for all the metrics of Fig. 2, our proposed method obtains the best results, which are coherent to the visual effects. In Table II, CNN also achieves the best performance except for SSIM. One possible reason for this small disturbance in SSIM is that most part of the abdomen image is soft tissue and BM3D is more preferable to this situation.

Table III shows the average values of the measurements for all the 100 images in the testing set. It can be observed that all the metrics demonstrate the best performance of the proposed CNN based method.

Table III Quantitative measurements (average values) associated with different algorithms for the images in testing set

|  | Low-dose | TV | KSVD | BM3D | CNN200 |
|---|---|---|---|---|---|
| PSNR | 36.3975 | 41.5021 | 40.8445 | 41.5358 | **42.1514** |
| RMSE | 0.0158 | 0.0087 | 0.0096 | 0.0088 | **0.0080** |
| SSIM | 0.8644 | 0.9447 | 0.9509 | 0.9610 | **0.9707** |

### 3.3. Qualitative measurement

For qualitative evaluation, 20 normal-dose images from the testing set and their corresponding low-dose versions, and 20 processed images from different methods are randomly selected for consideration. Artifact reduction, noise suppression, contrast retention and overall image quality are included as qualitative indicators with five grade assessment (1=worst and 5=best). Three radiologists (R1-R3) with more than 8 years of clinical experience evaluated these images and provided their scores. The low-dose images and the processed images with different algorithms are compared with the original normal-dose images and the student t test with $P < 0.05$ regarded as significantly different are performed to valid the discrepancy. The statistical results are illustrated in Table IV.

As demonstrated in Table IV, the qualities of low-dose images are much lower than normal-dose ones for all the scores. All the methods significantly improve the image quality in all qualitative indices and BM3D and the proposed CNN based algorithm achieve

Table IV Summary of statistical analysis of image quality scores for different algorithms (mean±SD)

|  |  | Normal-dose | Low-dose | TV | KSVD | BM3D | CNN200 |
|---|---|---|---|---|---|---|---|
| Artifact reduction | R1 | 3.95±0.60 | 1.95±0.69* | 2.85±0.59* | 2.60±0.50* | 3.45±0.60* | 3.65±0.58 |
|  | R2 | 4.05±0.51 | 1.95±0.51* | 2.95±0.39* | 2.90±0.39* | 3.40±0.60* | 3.80±0.52 |
|  | R3 | 3.95±0.22 | 1.75±0.44* | 2.80±0.41* | 2.75±0.44* | 3.40±0.50* | 3.80±0.41 |
|  | Average | 3.98±0.47 | 1.88±0.56* | 2.87±0.47* | 2.80±0.48* | 3.42±0.56* | 3.75±0.51 |
| Noise suppression | R1 | 3.90±0.64 | 1.70±0.57* | 3.05±0.51* | 3.00±0.46* | 3.60±0.50 | 3.65±0.49 |
|  | R2 | 4.25±0.44 | 1.85±0.49* | 3.50±0.51* | 3.40±0.50* | 4.05±0.39 | 4.05±0.51 |
|  | R3 | 3.90±0.31 | 1.55±0.51* | 3.00±0.56* | 3.05±0.60* | 3.65±0.49 | 3.75±0.55 |
|  | Average | 4.01±0.50 | 1.70±0.53* | 3.18±0.57* | 3.15±0.55* | 3.77±0.50* | 3.82±0.54 |
| Contrast retention | R1 | 3.85±0.59 | 1.65±0.49* | 2.65±0.49* | 2.80±0.52* | 3.50±0.51* | 3.55±0.51* |
|  | R2 | 3.95±0.39 | 1.55±0.51* | 2.60±0.50* | 2.85±0.49* | 3.50±0.51* | 3.55±0.51* |
|  | R3 | 3.95±0.39 | 1.85±0.49* | 2.60±0.50* | 2.95±0.22* | 3.55±0.51* | 3.50±0.51* |
|  | Average | 3.92±0.46 | 1.68±0.50* | 2.62±0.49* | 2.87±0.43* | 3.52±0.50* | 3.53±0.50* |
| Overall image quality | R1 | 3.85±0.49 | 1.75±0.44* | 2.65±0.49* | 2.70±0.47* | 3.55±0.51 | 3.65±0.49 |
|  | R2 | 3.90±0.31 | 1.65±0.49* | 2.95±0.22* | 3.05±0.22* | 3.60±0.60* | 3.70±0.57 |
|  | R3 | 3.85±0.37 | 1.55±0.51* | 2.70±0.47* | 2.75±0.44* | 3.60±0.50 | 3.65±0.49 |
|  | Average | 3.87±0.39 | 1.65±0.48* | 2.7±0.43* | 2.83±0.42* | 3.58±0.53* | 3.67±0.51 |

* indicates $P < 0.05$, which means significantly different.

the best results. The scores for the CNN are closer to the ones of normal-dose and the t test show the similar trend that the differences between the normal-dose images and the results from CNN based method is not statistically significant in all the qualitative indices except contract retention.

**3.4 Sensitivity analysis**

In this section, two important factors, the noise level for training set and the size of training set are evaluated for their impacts on the performance of CNN.

*3.4.1 Noise level for training and testing set*

It is well known that the performance of learning based methods is related to the samples involved in the training set. The noise levels for the training and testing set are consistent in our former experiment. The results have proved that the effectiveness of the proposed CNN based method comparing to the state-of-the-art methods in a certain noise level ($b_0 = 10^5$). However, the noise level for the target image is not always available, so different combinations for noise levels (both training and testing set) are used to validate the robustness of our method:

(a) Training set: $b_0 = 10^5$ and testing set: $b_0 = 10^5$;

(b) Training set: $b_0 = 10^5$ and testing set: $b_0 = 5 \times 10^5$;

(c) Training set: $b_0 = 10^5$ and testing set: $b_0 = 5 \times 10^4$;

(d) Training set: $b_0 = 10^5$ and testing set: $b_0 = 3 \times 10^4$;

(e) Training set: mixed data with $b_0 = 5 \times 10^5$, $10^5$, $5 \times 10^4$ and $3 \times 10^4$ and testing set: $b_0 = 5 \times 10^5$;

(f) Training set: mixed data with $b_0 = 5 \times 10^5$, $10^5$, $5 \times 10^4$ and $3 \times 10^4$ and testing set: $b_0 = 10^5$;

(g) Training set: mixed data with $b_0 = 5 \times 10^5$, $10^5$, $5 \times 10^4$ and $3 \times 10^4$ and testing set: $b_0 = 5 \times 10^4$;

(h) Training set: mixed data with $b_0 = 5 \times 10^5$, $10^5$, $5 \times 10^4$ and $3 \times 10^4$ and testing set: $b_0 = 3 \times 10^4$.

The quantitative results are given in Table V. CNN200-1 denotes the training set only includes single type data (situations (a)-(d)) and CNN200-4 denotes the training set includes mixed data (situations (e)-(h)). In all the cases except BM3D-F, the parameters in ASD-POCS, KSVD and BM3D are adjusted to achieve best average PSNR values for different noise levels. BM3D-F indicates running BM3D with fixed parameter for $b_0 = 10^5$. From Table V, two observations can be obtained: (1) CNN200-4 with mixed training set has best results in almost every metric in all the situations with different noise levels. Especially at the high noise levels, the advantage for CNN200-4 is more obvious. (2) When the noise level is low ($b_0 = 5 \times 10^5$), BM3D achieves best performance than other methods. As the noise level increases, CNN200-1 obtains competitive results than BM3D, which can be seen an evidence for the robustness at different noise levels. Meanwhile, we must mention here that the parameters in other methods are adjusted for different noise levels and it can be predicted that if the parameters are fixed, the performance for other methods will deteriorate for other noise levels. BM3D-F gives an example for this situation. It can be seen that the results for other noise levels (except $b_0 = 5 \times 10^5$) are worse than BM3D. The CNN200-1 will have even slightly better results than BM3D-F when the noise level is blind.

Table V. Quantitative measurements (average values) associated with different algorithms for different combinations of training and testing set

| Noise level of testing data | | Low-dose | TV | KSVD | BM3D | BM3D-F | CNN200-1 | CNN200-4 |
|---|---|---|---|---|---|---|---|---|
| $b_0 = 5 \times 10^5$ | PSNR | 41.0481 | 44.8030 | 44.0567 | 44.2798 | 43.0424 | 43.4386 | **43.6420** |
| | RMSE | 0.0091 | **0.0061** | 0.0064 | 0.0063 | 0.0072 | 0.0069 | 0.0067 |
| | SSIM | 0.9478 | 0.9735 | 0.9778 | 0.9796 | 0.9735 | 0.9766 | **0.9793** |
| $b_0 = 10^5$ | PSNR | 36.3975 | 41.5021 | 40.8445 | 41.5358 | 41.5358 | 42.1514 | **42.2282** |
| | RMSE | 0.0158 | 0.0087 | 0.0096 | 0.0088 | 0.0088 | 0.0080 | **0.0079** |
| | SSIM | 0.8644 | 0.9447 | 0.9509 | 0.9610 | 0.9610 | 0.9707 | **0.9720** |
| $b_0 = 5 \times 10^4$ | PSNR | 33.8300 | 39.7729 | 38.9090 | 39.8928 | 39.1694 | 40.6509 | **41.1100** |
| | RMSE | 0.0214 | 0.0106 | 0.0121 | 0.0106 | 0.0123 | 0.0096 | **0.0091** |
| | SSIM | 0.7950 | 0.9221 | 0.9296 | 0.9492 | 0.9269 | 0.9579 | **0.9650** |
| $b_0 = 3 \times 10^4$ | PSNR | 31.8020 | 38.3796 | 37.3982 | 38.6025 | 36.5157 | 38.9457 | **40.0148** |
| | RMSE | 0.0272 | 0.0125 | 0.0144 | 0.0125 | 0.0172 | 0.0120 | **0.0103** |
| | SSIM | 0.7301 | 0.9001 | 0.9128 | 0.9369 | 0.8788 | 0.9356 | **0.9565** |

*3.4.2 Size of training set*

The size of training set is another factor we considered to validate the impact to the performance of CNN based methods. In this subsection, we extend the size of training set to 2000 and the testing set is same. The quantitative results are given in Table VI.

Table VI. Quantitative results for CNN based methods with different sizes of training set

|      | CNN200-1 | CNN200-4 | CNN2000-1 | CNN2000-4 |
|------|----------|----------|-----------|-----------|
| PSNR | 42.1514  | 42.2282  | 42.3507   | **42.4215** |
| RMSE | 0.0080   | 0.0079   | 0.0078    | **0.0078** |
| SSIM | 0.9709   | 0.9720   | 0.9717    | **0.9724** |

From Table VI, improvements of quantitative metrics can be seen with larger training set. However, increasement of the training samples will significantly boost the training time. How to balance the computational cost and the performance is an important issue should be carefully considered in the practical applications.

*3.4.3 Real data test*

In the former sections, the corresponding low-dose CT images in both training and testing set were generated by numerical simulations. To validate the effectiveness of our method for real data, low-dose raw projections from sheep lung perfusion were used. In this study, an anesthetized sheep was scanned at normal and low doses respectively on a SIEMENS Somatom Sensation 64-slice CT scanner (Siemens Healthcare, Forchheim, Germany) in a helical scanning mode. The normal-dose scan was acquired with 100 kVp, 150 mAs protocol and the low-dose scan was acquired with 80 kVp, 17 mAs protocol. All the sinograms of the central slice were extracted, which were in a fan-beam geometry. The radius of the trajectory was 57 cm. 1160 projections were uniformly collected over 360 degrees. For each projection, 672 detector bins were equi-angularly distributed and a field-of-view (FOV) of 25.05 cm in radius was defined. In this experiment, the reconstructed images were 768×768 pixels with a physical size of 43.63×43.63 cm.

The results from the low-dose sinogram are demonstrated in Fig. 4. It can be observed that there is obvious noise in the FBP reconstruction image in Fig. 4(b) and the steak artifacts exits near the high attenuation structures. ASD-POCS can remove most of the noise and artifacts, but blocky effect also raises and blurs some important structures. BM3D and CNN based method suppressed the noise better than KSVD, but steak artifacts still exit in both Fig. 4(d) and (e). Three regions indicated by red arrows give some examples where the CNN based method achieved best results. Fig. 5 shows the differences between low-dose FBP image and other methods, which can be seen as supplement evident for the ability of structure preservation. Only noise and artifacts can be seen in Fig. 5(d) which implies little structure information was lost.

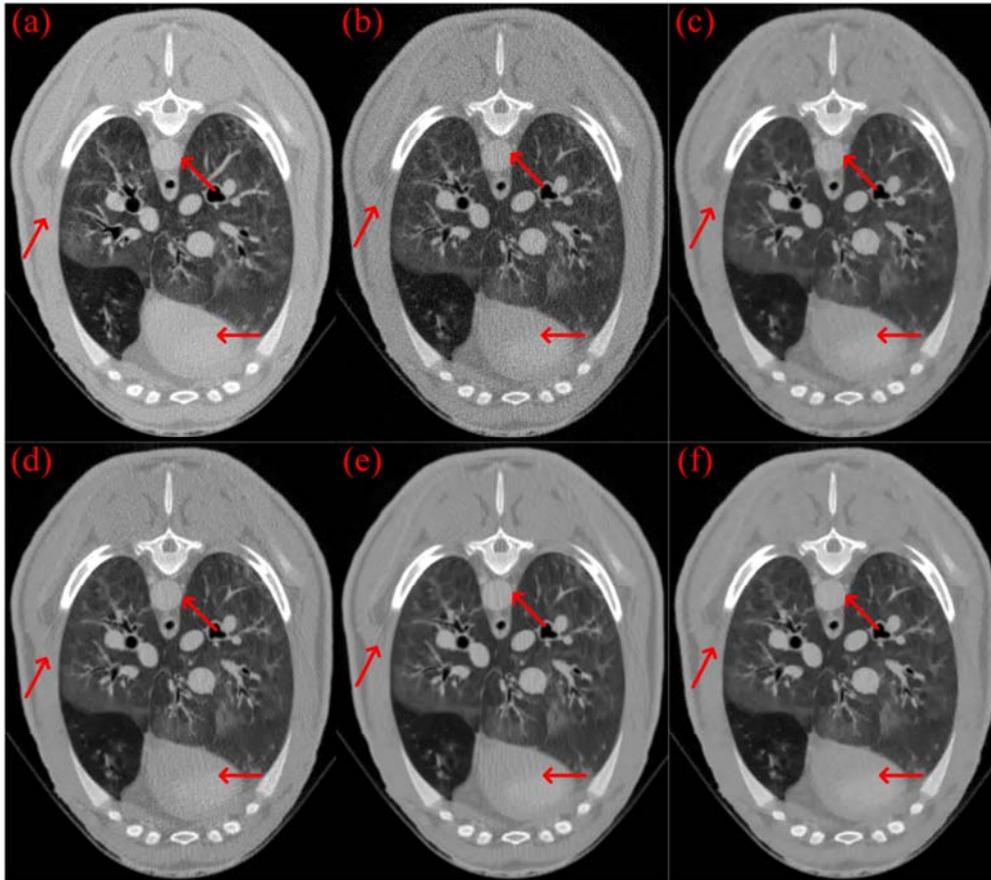

Fig. 4. Results images from a low-dose sinogram collected in the sheep lung CT. (a) Original normal-dose image; (b) the low-dose images; (c) the ASD-POCS reconstructed image; (d) the KSVD processed low-dose image; (e) the BM3D processed low-dose image; (f) the CNN processed low-dose image.

## 4. Discussion and conclusion

Deep learning has achieved exciting results in several fields in computer vision and image processing, such as image segmentation, object detection, object tracking, image supper-resolution, etc. In medical imaging, deep learning is only applied in medical image segmentation and lesion detection and its potential for other applications has not been explored. In this paper, a framework of deep convolution neural network is presented for noise reduction for low-dose CT. The proposed method learns a feature mapping from the low- to normal-dose images. After training stage, our method has achieved competitive performance than the state-of-the-art methods in both qualitative and quantitative evaluations. We also assess several factors for model performance tradeoffs, including the robustness of the proposed method for different noise levels, the training strategy and the size of training set. It also can be conjectured that larger capacity of the network may further improve the performance and we will investigate it in the future.

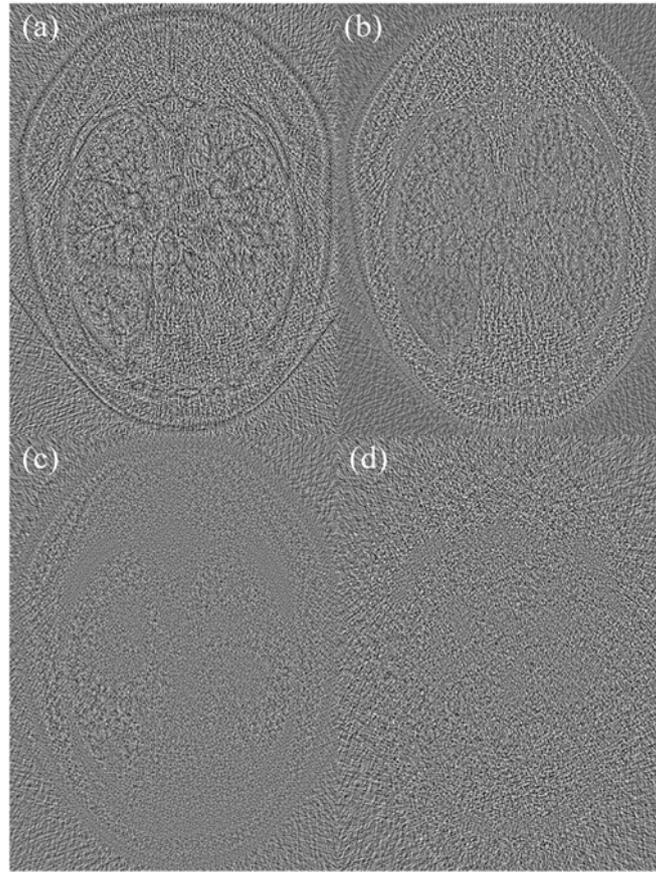

Fig. 5. The difference images between the low-dose FBP image and the results by (a) ASD-POCS, (b) KSVD, (c) BM3D, and (d) CNN.

Computational cost for the proposed CNN based method is concerned by the users. The main cost of computational time is spent on the training stage. Although the training is implemented with GPU mode, it is still time-consuming. For the smaller training set with 200 images, about $10^6$ patches were involved and it toke 17 hours. For the larger training set with 2000 images, about 107 patches were involved and 56 hours were spent. Other iterative methods do not have training phase, but the execution time is much longer than CNN based method. The average execution times (CPU mode) for ASD-POCS, KSVD, BM3D and CNN are 23.79, 40.88, 3.33 and 2.05 seconds. Actually, once the network is trained offline, CNN based method is more efficient than other methods in terms of real execution time.

In this study, we explore the performance for applying deep convolutional neural network to noise reduction in low-dose CT. The results demonstrate the potentials of CNN based method for medical imaging. In the future work, the proposed network structure can be applied to other topics in CT imaging, such as low-dose CT reconstruction, metal artifact reduction and interior CT. Another possible direction is to investigate other network architecture to deal with these problems.

**Reference**

[1] A. B. de Gonzalez and S. Darby, "Risk of cancer from diagnostic X-rays: Estimates for the UK and 14 other countries," Lancet, vol. 363, pp. 345–351, 2004.

[2] D. J. Brenner and E. J. Hall, "Computed tomography–An increasing source of radiation exposure," New Eng. J. Med., vol. 357, pp. 2277–2284, 2007.

[3] M. Balda, J. Hornegger and B. Heismann, "Ray contribution masks for structure adaptive sinogram filtering," IEEE Trans. Med. Imaging 30(5), 1116–1128 (2011).

[4] A. Manduca, L. Yu, J. D. Trzasko, et al., "Projection space denoising with bilateral filtering and CT noise modeling for dose reduction in CT," Med. Phys. 36(11). 4911-4919 (2009)

[5] T. Li, X. Li, J. Wang, et al., "Nonlinear sinogram smoothing for low-dose x-ray CT," IEEE Trans. Nucl. Sci. 51(5), 2505-2513 (2004).

[6] J. Wang, H. Lu, J. Wen and Z. Liang, "Multiscale penalized weighted least-squares sinogram restoration for low-dose x-ray computed tomography," IEEE Trans. Biomed. Eng. 55(3) 1022–1031 (2008).

[7] S. Tang and X. Tang, "Statistical CT noise reduction with multiscale decomposition and penalized weighted least squares in the projection domain," Med. Phys. 39(9) 5498-5512 (2012).

[8] E. Y. Sidky and X. Pan, "Image reconstruction in circular cone-beam computed tomography by constrained, total-variation minimization," Phys. Med. Biol. 53(17), 4777–4807 (2008).

[9] Y. Zhang, W. Zhang, Y. Lei, and J. Zhou, "Few-view image reconstruction with fractional-order total variation," J. Opt. Soc. Am. A 31(5), 981–995 (2014).

[10] Y. Zhang, Y, Wang, W. Zhang, et al., "Statistical iterative reconstruction using adaptive fractional order regularization," Biomed. Opt. Express 7(3), 1015-1029 (2016).

[11] Y. Zhang, W.-H. Zhang, H. Chen, M.-L. Yang, T.-Y. Li, and J.-L. Zhou, "Few-view image reconstruction combining total variation and a high-order norm," Int. J. Imaging Syst. Technol. 23(3), 249–255 (2013).

[12] Y. Chen, D. Gao, C. Nie, et al., "Bayesian statistical reconstruction for low-dose x-ray computed tomography using an adaptive weighting nonlocal prior," Comput. Med. Imaging Graph. 33(7) 495–500 (2009).

[13] J. Ma, H. Zhang, Y. Gao et al., "Iterative image reconstruction for cerebral perfusion CT using a pre-contrast scan induced edge-preserving prior," Phys. Med. Biol. 57(22), 7519-7542 (2012).

[14] Q. Xu, H. Yu, X. Mou, et al., "Low-dose x-ray CT reconstruction via dictionary learning," IEEE Trans. Med. Imaging 31(9), 1682–1697 (2012).

[15] J.-F. Cai, X. Jia, H. Gao, et al., "Cine cone beam CT reconstruction using low-rank matrix factorization: algorithm and a proof-of-principle study," IEEE Trans. Med. Imaging 33(8), 1581-1591 (2014).

[16] Y. Chen, Z. Yang, Y, Hu et al., "Thoracic low-dose CT image processing using an artifact suppressed large-scale nonlocal means," Phys. Med. Biol. 57(9) 2667-2688 (2012).

[17] J. Ma, J. Huang, Q. Feng, et al., "Low-dose computed tomography image restoration using previous normal-dose scan," Med. Phys. 38(10), 5713-5731 (2011).

[18] Z. Li, L. Yu, J. D. Trzasko, et al., "Adaptive nonlocal means filtering based on local noise level for CT denoising," Med. Phys. 41(1), 011908 (2014).

[19] M. Aharon, M. Elad, and A. Bruckstein, "K-SVD: An algorithm for designing overcomplete dictionaries for sparse representation," IEEE Trans. Signal Process. 54(11), 4311-322 (2006).

[20] Y. Chen, X. Yin, L. Shi, et al., "Improving abdomen tumor low-dose CT images using a fast dictionary learning based processing," Phys. Med. Biol. 58(16), 5803–5820 (2013).

[21] P. F. Feruglio, C. Vinegoni, J. Gros, et al., "Block matching 3D random noise filtering for absorption optical projection tomography," Phys. Med. Biol. 55(18), 5401-5415 (2010).

[22] K. Sheng, S. Gou, J. Wu, et al., "Denoised and texture enhanced MVCT to improve soft tissue conspicuity," Med. Phys. 41(10) 101916 (2014).

[23] D. Kang, P. Slomka, R. Nakazato, et al., "Image denoising of low-radiation dose coronary CT angiography by an adaptive block-matching 3D algorithm," Proc. SPIE 8669, Medical Imaging 2013: Image Processing, 86692G (2013).

[24] G. E. Hinton, S. Osindero, and Y.-W. Teh, "A fast learning algorithm for deep belief nets," Neural Comp. 18(7) 1527-1554 (2006).

[25] G. E. Hinton and R. Salakhutdinov, "Reducing the dimensionality of data with neural networks," Science 313(5786) 504-507 (2006).

[26] Y. LeCun, Y, Bengio and G. Hinton, "Deep learning," Nature 521(7553) 436-444 (2015).

[27] V. Jain and H. Seung, "Natural image denoising with convolutional networks," in Proc. Adv. Neural Inf. Process. Syst. 769-776 (2008).

[28] C. Dong, C. C. Loy, K. He, et al., "Image super-resolution using deep convolutional networks," IEEE Trans. Pattern Anal. Mach. Intell. 38(2) 295-307 (2016).

[29] L. Xu, J. S. J. Ren, C. Liu, et al., "Deep convolutional neural network for image deconvolution," in Proc. Adv. Neural Inf. Process. Syst. 1790-1798 (2014).

[30] J. Xie, L. Xun and E. Chen, "Image denoising and inpainting with deep neural networks," in Proc. Adv. Neural Inf. Process. Syst. 350-352 (2012).

[31] H. C. Burger, C. J. Schuler and S. Harmeling, "Image denoising: can plain neural networks compete with BM3D?" In Proceedings of the IEEE Conference on Computer Vision and Pattern Recognition (CVPR), 2392-2399 (2012).

[32] R. Wang and D. Tao, "Non-local auto-encoder with collaborative stabilization for image restoration," IEEE Trans. Image Process. 25(5) 2117-2129 (2016).

[33] F. Agostinelli, M. R, Anderson and H. Lee, "Adaptive multi-column deep neural networks with application to robust image denoising," in Proc. Adv. Neural Inf. Process. Syst. 1493-1501 (2013).

[34] S. Liao, Y. Gao, A. Oto, et al., "Representation learning: a unified deep learning framework for automatic prostate MR segmentation," Med. Image Comput. Comput. Assist. Interv. 16(2) 254-261 (2013).

[35] K. H. Cha, L. Hadjiiski, R. K. Samala, et al., "Urinary bladder segmentation in CT urography using deep-learning convolutional neural network and level sets," Med. Phys. 43(4) 1882-1896 (2016).

[36] M. Kallenberg, K. Petersen, M. Nielsen, et al., "Unsupervised deep learning applied to breast density segmentation and mammographic risk scoring," IEEE Trans. Med. Imaging 35(5) 1322-1331 (2016).

[37] J. Xu, L. Xiang, Q. Liu, et al., "Stacked sparse autoencoder (SSDA) for nuclei detection on breast cancer histopathology images," IEEE Trans. Med. Imaging 35(1) 119-130 (2016).

[38] K. Sirinukunwattana, S. H. A. Raza, Y.-W. Tsang, et al., "Locality sensitive deep learning for detectionand classification of nuclei in routine colon cancer histology images," IEEE Trans. Med. Imaging 35(5) 1196-1206 (2016).

[39] H. C. Shin, M. R. Orton, D. J. Collins, et al., "Stacked autoencoders for unsupervised feature learning and multiple organ detection in a pilot study using 4D patient data," IEEE Trans. Pattern Anal. Mach. Intell. 35(8) 1930-1943 (2016).

[40] S. Wang, Z. Su, L. Ying, et al., "Accelerating magnetic resonance imaging via deep learning,"2016 IEEE 13th International Symposium on Biomedical Imaging (ISBI), 514-517 (2016).

[41] H, Zhang, L. Li, K. Qiao, et al., "Image predication for limited-angle tomography via deep learning with convolutional neural network," arXiv:1607.08707 (2016).

[42] G. Wang, "A perspective on deep imaging," arXiv:1609.04375 (2016).

[43] E. J. Candès, J. Romberg, and T. Tao, "Robust uncertainty principles: Exact signal reconstruction from highly incomplete frequency information," IEEE Trans. Inf. Theory 52(2), 489-509 (2006).

[44] M. Elad and M. Aharon, "Image denoising via sparse and redundant representations over learned dictionaries," IEEE Trans. Image Process. 15(2), 3736-3745 (2006).

[45] V. Nair and G. E. Hinton, "Rectified linear units improve restricted Boltzmann machines," in Proc. Int. Conf. Mach. Learn., 807-814 (2010).

[46] Y. LeCun, L. Bottou, Y. Bengio, and P. Haffner, "Gradient-based learning applied to document recognition," Proc. IEEE, 86(11), 2278-2324 (1998).

[47] Z. Wang, A. C. Bovik, H. R. Sheikh, and E. P. Simoncelli, "Image quality assessment: from error visibility to structural similarity," IEEE Trans. Image Process. 13(4), 600–612 (2004).

[48] R. L. Siddon, "Fast calculation of the exact radiological path for a three-dimensional CT array," Med. Phys. 12(2), 252-255 (1985).